\documentclass{article}

\usepackage{graphicx}  
\usepackage{amsmath}   
\usepackage{amssymb}   

\def\eq#1{{Eq.~(\ref{#1})}}

\usepackage{bm}
\usepackage{latexsym}
\usepackage{dcolumn}
\usepackage{amsmath,amsfonts,amssymb}
\usepackage{graphicx,epsfig}
\usepackage{color}
\usepackage{enumitem}
\usepackage[affil-it]{authblk}

\def \lp { L_0 }

\def\nn{\nonumber}

\def\l{\left}
\def\r{\right}

\reversemarginpar

\def \mA {A\l[\sigma;\lp \r]}

\begin{document}

  \title{Entropy density of spacetime from the zero point length}

\author{Dawood Kothawala%
  \thanks{Electronic address: \texttt{dawood@physics.iitm.ac.in}}}
\affil{Department of Physics, IIT Madras, Chennai - 600 036, India}

\author{T. Padmanabhan%
  \thanks{Electronic address: \texttt{paddy@iucaa.ernet.in}}}
\affil{IUCAA,  Post Bag 4, Ganeshkhind, Pune - 411 007, India}

\date{Dated: \today}

  
 \maketitle
  
  \begin{abstract}
It is possible to obtain the gravitational field equations in a large class of theories from a  thermodynamic variational principle which uses the gravitational heat density $\mathcal{S}_g$ associated with null surfaces. This heat density is related to the discreteness of spacetime at Planck scale, $L_P^2 = (G\hbar / c^3)$, which assigns  $A_{\perp}/L_P^2$ degrees of freedom to any area $A_{\perp}$. On the other hand, it is also known that the surface term $K\sqrt{h}$ in the gravitational action principle correctly reproduces the heat density of the null surfaces. We provide a link between these ideas by obtaining  $\mathcal{S}_g$, used in emergent gravity paradigm, from the surface term in the action in Einstein's gravity. This is done using the notion of a nonlocal qmetric -- introduced recently [arXiv:1307.5618, 1405.4967] -- which allows us to study the effects of {\it zero-point-length} of spacetime at  the  transition scale between quantum and classical gravity. Computing $K\sqrt{h}$ for the qmetric in the appropriate limit directly reproduces the entropy density  $\mathcal{S}_g$ used in the emergent gravity paradigm. 
\end{abstract}

 
The thermodynamic  potentials like entropy density ($s$), the heat density ($Ts$), the free energy density $(\rho-Ts)$ etc. provide a link between the microscopic dynamics of molecules and macroscopic dynamics described in terms of standard thermodynamic variables like pressure, temperature etc. Recent work has shown that the field equations of gravity, describing the evolution of spacetime, are akin to the equations describing, say, the gas dynamics \cite{tpreviews, grtp}. These field equations, for a  large class of theories of gravity, can be obtained \cite{entropy-functional,llreview} by extremising the total heat density  $\mathcal{S}=\mathcal{S}_g+\mathcal{S}_m$ where 
 $\mathcal{S}_m$ is the matter heat density and  $\mathcal{S}_g[n]$ the gravitational heat density. The latter depends on a  vector field $n^i$ of constant norm and is given by \cite{grtp} 
\begin{equation}
 \mathcal{S}\propto [(\nabla_in^i)^2-\nabla_i n^j \nabla_jn^i]=R_{ab} n^an^b +(\text{tot.  div.})
\label{ts}
\end{equation}
in the case of Einstein's gravity.
 Extremising $\mathcal{S}$ with respect to all vector fields $n^i$ simultaneously, leads to a constraint on the background metric which turns out to be identical to the field equations. 
 
 If the ideas of emergent gravity paradigm are correct, we should be able to obtain this expression from a more microscopic approach. We will show how this can be done.
 
There are three facts which guide us in this task which we will first describe.
 \begin{enumerate}
  \item 
 It seems inevitable that that the existence of some `atoms of spacetime' is related to an effective \textit{discreteness} at Planck scale ($L_P^2 = (G\hbar / c^3)$), which allows us to assign,  $N_{\rm sur}=A_{\perp}/L_P^2$ degrees of freedom with any area $A_{\perp}$. In fact one can show that \cite{grtp} the time evolution of geometry 
in a 3-volume is driven by the difference $[N_{\rm sur}-N_{bulk}]$ where $N_{bulk}$ is the number of bulk degrees of freedom in the volume and 
 $N_{\rm sur}$ is the number of surface degrees of freedom in the boundary. (Observers who perceive a time-independent metric will also note that spacetime exhibits holographic equipartition in the sense of $N_{\rm sur}=N_{bulk}$.)  So, clearly, we need to incorporate the notion of `zero-point-area' $L_P^2$ in a suitable manner if we hope to obtain $\mathcal{S}_g $ from a more microscopic description.
 \item
 We know from standard discussion of horizon thermodynamics that the surface term $\mathcal{A}_{\rm sur}$ in the gravitational action (defined as the integral of $K\sqrt{h}$), is closely related to the entropy and heat densities. More precisely, the surface Hamiltonian 
 \begin{equation}
  H_{\rm sur}\equiv\frac{\partial \mathcal{A}_{\rm sur}}{\partial t}=\frac{\partial }{\partial t}\left[\frac{1}{8\pi  L_P^2}\int_{\mathcal{H}} K\sqrt{h}d^3x\right]
  \end{equation} 
when evaluated on a local Rindler horizon $\mathcal{H}$ with surface gravity $\kappa$ and transverse area $A_\perp$, gives \cite{bibastp}
the heat content(which is the same as enthalpy in this context):
\begin{equation}
H_{\rm sur}\to \frac{\kappa A_\perp}{8\pi  L_P^2}=TS                                             
\end{equation} 
 If we now perform 
 the Euclidean continuation in $t$, then the natural range of integration for the Euclidean time $t_E$ is $0<t_E<(2\pi/\kappa)$. This will give the result:
 \begin{eqnarray}
  \mathcal{A}_{\rm sur}^E &=& \int_{\mathcal{H}} d^3x K\sqrt{h} 
  \nn \\
  &=& \frac{2\pi}{\kappa}\times\left(\frac{\kappa A_\perp}{8\pi  L_P^2} \right)=\frac{A_\perp}{4L_P^2}=S
  \label{entro}
 \end{eqnarray} 
 showing that the Euclidean surface action is the entropy. Therefore, the entropy \textit{density} of spacetime, when evaluated around any event after Euclidean continuation, is essentially $(K\sqrt{h})$.
 
 \item The above two facts suggest that one should be able to obtain the entropy density used in the emergent gravity paradigm, given in \eq{ts}, from $K\sqrt{h}$ in a suitable limit.  The operational difficulty in this program, of course, is the following: Quantities like $K\sqrt{h}$ are well defined on a differentiable manifold with a metric, normal vectors etc. But the entropy arising from  $A_\perp/L_P^2$  degrees of freedom associated with an area $A_\perp$, requires incorporation of the zero point area into the spacetime which cannot be done without modifying the usual, local, description of spacetime. We need a suitable prescription which incorporates the quantum gravitational effects (in particular the existence of zero point area $L_P^2$), at scales reasonably bigger than $L_P^2$ but not totally classical.
 So we need a notion of an ``effective'' metric $q_{ab}$ in a spacetime (with a classical metric $g_{ab}$) such that it can incorporate the  effects of the zero point area $L_P^2$. We can then compute $K\sqrt{h}$ for this effective metric. In the appropriate limit, this should give us the entropy density of the spacetime and --- if our ideas are correct --- the resulting entropy density should match with the one in \eq{ts}.

 \end{enumerate}

 Fortunately, the key last step of introducing an effective metric $q_{ab}$ with the necessary properties has already been achieved.
 We recently described in Ref. (\cite{dk-ml, cheshire}) the notion of a qmetric which is capable of doing precisely this, which we shall briefly recall:

In a classical spacetime one can introduce a geodesic interval $\sigma^2(P,p)$ between any two events $P$ and $p$ which contains the same amount of information as the metric tensor $g_{ab}$.
The key difference, of course, is that $\sigma^2(P,p)$ is a biscalar (and hence nonlocal) while the metric is local. Various geometric quantities at $P$ can be constructed by taking suitable derivatives of $\sigma^2(P,p)$ with respect to the coordinate $p$ and then taking the limit $p\to P$. (More details can be found in \cite{poisson-lrr, cheshire}.) At the classical level, the geometry can be characterized either by $g_{ab}$ or by $\sigma^2(P,p)$.

When we try to incorporate the effects of quantum gravity, there is an advantage in starting from a description in terms of $\sigma^2(P,p)$ rather than from the metric. This is because, while we have no universal rule to understand 
how quantum gravity modifies the metric,  there is considerable amount of evidence (see e.g., \cite{zpl}) which suggests that $\sigma^2(P,p)$ is modified by
\begin{equation}
 \sigma^2 \to \sigma^2 + \lp^2; \qquad \lp^2=\mu^2 L_P^2
\label{funda}
\end{equation}
where $\mu$ is a factor of order unity \cite{comment}. That is, one can capture the lowest order quantum gravitational effects by introducing a zero point length in spacetime along the lines suggested by \eq{funda}. Once we accept this, we can introduce a second rank symmetric \textit{bitensor} $q_{ab}(p,P)$, constructed such that it will lead to the geodesic interval $\sigma^2 + L_0^2$ if the original metric had the geodesic interval $\sigma^2$. This is done in \cite{dk-ml, cheshire} by associating with  a metric $g_{ab}$ (which has the corresponding geodesic interval $\sigma^2$) a \textit{nonlocal} symmetric bitensor $q_{ab}(p,P)$ by the relation:  
 \begin{equation}
q_{ab}(p,P; \lp^2) = Ag_{ab} -    \l( A  - \frac{1}{A} \r) \; n_a n_b
\label{eq:key1}
\end{equation}
where $g_{ab} = g_{ab}(p)$  is the classical metric tensor, $\sigma^2 = \sigma^2 (p,P)$ is the corresponding classical geodesic interval  and 
\begin{equation}
\mA = 1 + \frac{\lp^2}{\sigma^2}; \qquad n_a = \frac{\nabla_a \sigma^2}{2 \sqrt{  \sigma^2}}
\label{n1}
\end{equation}
(The derivation of this form of qmetric and its properties are given in \cite{dk-ml, cheshire}).
Working with the qmetric we can capture some of the effects of quantum gravity --- especially those arising from the existence of the zero point area --- \textit{without leaving the comforts of the standard differential geometry.} 

There are several non-trivial effects arising from the nonlocal description of geometry in terms of the qmetric, discussed at length out in detail in \cite{cheshire}, with the key point 
being the following: Suppose $\phi(P|g)$ is some scalar computed from the metric $g_{ab}$ and its derivatives. (The $\phi$ could, for example, be $R$, or $R_{ab}R^{ab}$ etc.). When we carry out the corresponding algebra using $q_{ab}(p,P)$ (with all differentiations carried out at the event $p$) we will end up getting a nonlocal (biscalar) $\phi(p,P; \lp^2|q)$ which depends on two events $(p,P)$ and on $\lp^2$. To obtain a local result, we now take the limit of $\sigma\to0$ (that is, $p\to P$) keeping $L_0^2$ finite. The resulting $\phi(P,P; \lp^2|q)$ will show quantum gravitational residual effects due to nonzero $L_0^2$. 
The key features of this approach arise from the non-commutativity of the limits:
\begin{equation}
\lim_{L_0^2\to0}\lim_{\sigma^2\to0} \phi(p,P; \lp^2|q)\neq
\lim_{\sigma^2\to0}\lim_{L_0^2\to0}\phi(p,P; \lp^2|q)
\end{equation} 
The limit of the right hand side is trivial.
When we  take the limit of $L_0^2\to0$, keeping $\sigma^2$ finite, then $q_{ab}\to g_{ab}$ and all the derivatives of qmetric will coincide with the corresponding derivatives of the metric and $\phi(p,P; \lp^2|q)\to\phi(P|g)$. This arises because, in \eq{eq:key1}, only the combination $L_0^2/\sigma^2$ introduces non-trivial effects and this term and all derivatives vanish when $L_0^2\to0$. But when we take the limit of $\sigma\to0$ first keeping $L_0^2$ finite, the qmetric actually diverges. So we have no assurance that we will even get anything sensible when we take the limit; surprisingly, we do. This is how we obtain non-trivial effects. (For details, see \cite{cheshire}.)
The motivation, justification and several properties of the qmetric and the two limits were described in detail in \cite{cheshire} and will not be repeated here. 
 
After this preamble, we return to our main focus, the surface term $K\sqrt{h}$ in the gravitational action. Given a fixed spacetime event $P$, the most natural {\it surface} $\Sigma$ on which to evaluate this term is the one formed by events $p$ at a constant  geodesic interval $\sqrt{|\sigma^2(p,P)|}=\lambda$ from $P$. The intrinsic as well as extrinsic geometry of such a surface is completely determined by the geodesic structure of the background manifold, and hence is completely characterized by invariants built out of spacetime curvature. The mathematical expressions we shall need here can be found in \cite{cheshire}, and several additional geometrical aspects of such {\it equi-geodesic} surfaces are discussed in \cite{dk-ext-geom}. 

We will use the qmetric and compute $K\sqrt{h}$ and demonstrate that it does lead to the entropy density $\mathcal{S}_g$ in \eq{ts}. This is a relatively straightforward (though somewhat lengthy) computation and we shall describe the key steps. For clarity, we will work in a $D=4$ Euclidean space (the final result is same for Lorentzian signature), and use units with $L_P=1$ so that $L_0=\mu$. \textit{In the local Rindler frame around $P$, the origin of $t_E-x$ plane will be the horizon and hence the limit of $p\to P$ corresponds to computing a quantity on the horizon.} We want to compute $K\sqrt{h}(p,P,\mu^2)$ for the qmetric and take the limit $p\to P$ (i.e., $\lambda\to0$) to obtain the quantum corrected entropy density. 

The $[K\sqrt{h}]_q$ for the qmetric can be easily related to the corresponding quantity evaluated for the metric $g_{ab}$ by the relation
 \begin{equation}
 \left[K\sqrt{h}\right]_q = A^2\left[K\sqrt{h}\right]_g + \frac{3}{2} \sqrt{h} \nabla_{\bm n} A
  \label{zero}
 \end{equation} 
where $\nabla_{\bm n} \equiv n^i \nabla_i$. The extrinsic curvature tensor on this surface has a series expansion in $\lambda$ given by (see \cite{cheshire, dk-ext-geom})
 \begin{equation}
K = \frac{3}{\lambda} - \frac{1}{3} \lambda\ \mathcal{S}(P) + \mathcal{O}(\lambda^2)
  \label{one}
 \end{equation} 
where $\mathcal{S}(P) = R_{ab} n^a n^b|_P$. Since by definition $K=\partial(\ln \sqrt{h})/ \partial \lambda$, this leads to the following series expansion for $\sqrt{h}$:
\begin{equation}
  \sqrt{h} = \lambda^3 \left[ 1 - \frac{1}{6}  \mathcal{E}(P)\, \lambda^2 + \mathcal{O}(\lambda^3)\right]
  \label{two}
 \end{equation} 
In the units we are using $\sqrt{h}$ incorporates the length dimensions and $K\sqrt{h}$ has dimensions $[\rm length]^2$. We also have 
 \begin{equation}
 \nabla_{\bm q} A = -\ \frac{2 \mu^2}{\lambda^3}
 \label{three}
 \end{equation}
  Substituting \eq{one}, \eq{two} and \eq{three} in \eq{zero} we get the result
  \begin{equation}
 \left[K\sqrt{h}\right]_q = 3A \lambda^2 - \frac{5}{6} (A\lambda^4)\ R_{ab} n^an^b \left[ 1 + \frac{2}{5} \, \frac{\mu^2}{\lambda^2}\right] + \mathcal{O}(\lambda)
 \end{equation}
 Using $A= 1+(\mu^2/\lambda^2)$ and taking the coincidence limit $\lambda \to 0$, we get the final result
\begin{eqnarray}
\lim \limits_{\lambda \to 0} \left[K\sqrt{h}\right]_q &=& 3 \mu^2 - \frac{\mu^4}{3} \ R_{ab} n^an^b 
\nn \\
&=& \mathcal{S}_0 - \frac{\mu^4}{3} \mathcal{S}_g
\label{final1}
 \end{eqnarray}
with all quantities on RHS now evaluated at $P$. 
The term $\mathcal{S}_0 = 3\mu^2$ can be thought of as the zero point entropy density of the spacetime which is a new feature. Its numerical value depends on the ratio $\mu=L_0/L_P$ which we expect to be of order unity and we will comment on it towards the end. The second term is exactly the heat density used in emergent gravity paradigm.
 
This result is significant in several ways which we shall now describe.
 
 The most important feature of our result is that it reproduces correctly (except for an unimportant multiplicative constant) the entropy density $\mathcal{S}_g \propto R_{ab} n^a n^b$ used in emergent gravity paradigm.  \textit{This tells us that the entire program has a remarkable level of internal consistency.} There is no way one could have guessed this result a priori and, in fact, there is no assurance that the result should even be finite in the coincidence limit of $\sigma^2 \to 0$. The qmetric itself diverges when $\sigma^2 \to 0$ and its derivatives diverge faster.  It is a nice and a non-trivial feature that the final result is free of any divergence. 
 
 Second,  it is rather satisfying to obtain this result from $K\sqrt{h}$ part of the action rather than from the $R\sqrt{-g}$ part of the action. It has been shown in several previous works \cite{ayantp} that there is an intimate relationship between the surface and bulk parts of the gravitational action and hence we would have expected the correct entropy density $\mathcal{S}_g$ to emerge from either of them if it emerges from one of them. This expectation is correct and indeed we have shown earlier \cite{cheshire} that a similar analysis with the bulk part of the action does lead to the correct entropy density. There are, however, some crucial differences between these two approaches. The computation in Ref. \cite{cheshire} leads to an expression with a divergent term which, however, can be regularized to give the correct final result. The computation here, starting from the surface term, however does not lead to any divergences and there is no need to regularize the final result.  This is a mathematically \textit{non-trivial} fact which arises from a \textit{delicate cancellation} of divergences\footnote{For example, if we attempted the same procedure with the bulk cosmological constant term in the action ($\Lambda\sqrt{-g}$) by converting it to ($\Lambda\sqrt{-q}$)
and taking the $\sigma^2\to0$ limit, it will diverge unless $\Lambda=0$. This result is probably telling us that a microscopic approach cannot accommodate a nonzero cosmological constant.} between the two terms on the right hand side of \eq{zero}. More specifically, the numerical factor and the structure of second of these terms depend on the (disformal) form of the qmetric, and an arbitrary, ad hoc deformation of geometry will \textit{not} lead to similar cancellation of divergences (see, however,  \cite{comment}). 
 
 Further, as we argued earlier, $K\sqrt{h}$ does have the natural interpretation of (being proportional to) the heat density on the horizon.  Note that when we work in the Euclideanized local Rindler frame around an event $P$, the Rindler horizon gets mapped to the origin of the $(x, t_E)$ plane. The coincidence limit of $p\to P$ is precisely the same as taking the horizon limit in the local Rindler frame. In this limit, as is obvious from \eq{entro}, $K\sqrt{h}/8\pi$ gives the entropy density. So if we had taken the limit $L_0\to0$  first (when $q_{ab}\to g_{ab}$ etc.) we would have recovered this standard result. 

Finally, the most intriguing feature of our result is the discovery of ``zero point entropy density'' represented by the first term
 $\mathcal{S}_0 = 3\mu^2$. Since this is an entropy density, it tells us that the total zero point entropy in a sphere of Planck radius is given by
 \begin{equation}
  S_0 = \frac{4\pi}{3} \times 3 \mu^2 = 4\pi\mu^2
  \label{fourpi}
 \end{equation} 
 Recently, it has been shown that  the cosmological constant problem can be solved within the emergent gravity paradigm if one could attribute a value $4\pi$ to the measure of degrees of freedom in the universe at Planck epoch, if the inflation took place at GUTs scale. This measure remains as a conserved quantity during the subsequent evolution and allows one to determine the numerical value of the cosmological constant (see, for details, Ref. \cite{hptp}). On the other hand if the inflation took place at Planck scales, we need $\mu^2\approx 1.2$ (see \cite{tpqg})
which is quite consistent with \eq{fourpi}.

Unfortunately, the value of $\mu$ cannot be determined from the analysis of pure gravity sector for two reasons. First of all, there can be a numerical factor multiplying $K\sqrt{h}$ to give the entropy density. In the standard approach, this term is $(1/8\pi L_P^2)K\sqrt{h}$ but it is not clear whether we should use the same expression in a microscopic theory. Second, the overall coupling between gravity and matter is undetermined until we have introduced the matter sector which we have not yet done. If we assume that the total heat density, maximized to get the field equations is the sum of gravitational and matter heat densities (with the latter being $\mathcal{S}_{m}=T_{ab}n^an^b$; see e.g. \cite{grtp,entropy-functional}), then one can determine the value of $\mu$. (Incidentally, the negative sign of the second term in \eq{final1} is important for the consistency of this result; the fact that it comes out right is another consistency check for this approach.) But it is possible for a microscopic approach to modify the matter sector term to $\mathcal{S}_{m}=\lambda T_{ab}n^an^b$ where $\lambda$ is a numerical factor. So, altogether there is a possibility of yet another undetermined numerical factor in the theory. To see its effect, let us take the gravitational entropy term as just $K\sqrt{h}$ and write the matter sector term as $\mathcal{S}_{m}=\lambda T_{ab}n^an^b$ where $\lambda$ is a numerical factor. Then simple algebra shows that, to reproduce Einstein's equations $G_{ab}=8\pi T_{ab}$ with correct coefficient, we need  $(\mu^4/24\pi\lambda)=(1/8\pi)$ or $\mu^2=\sqrt{3\lambda}$.  While this is in the right range to solve the cosmological constant problem, the numerical factor cannot be fixed until we have obtained the heat density of the matter sector from a similar description. But it is clear that the result in \eq{fourpi}, which brings in a zero-point-entropy density, could provide a more detailed and microscopic justification for this idea. This issue is under investigation.

\section*{Acknowledgments}
  The research work of TP is partially supported by the J.C.Bose Fellowship of DST, India. DK thanks IUCAA, Pune, where part of this work was done, for hospitality. We thank Bibhas Majhi for comments on the earlier version of the paper.




\begin{thebibliography}{99}
\bibitem{comment} {It is possible to consider the general case where $\sigma^2 \to S(\sigma^2)$ where the function $S$ will presumably arise from some fundamental framework 
for quantum gravity \cite{cheshire}. Surprisingly, our key result turns out to be independent of the precise form for $S(\sigma^2)$ as long as $S(0)=L_0^2$!}
\end{thebibliography}
 \end{document}